\documentclass[aps,prl,twocolumn,showpacs,superscriptaddress]{revtex4-1}
\usepackage{graphicx}
\usepackage{dcolumn}
\usepackage{bm}
\usepackage{amsmath}
\usepackage{color}
\usepackage[normalem]{ulem}

\begin{document}

\newcommand{\bra}[1]{\left\langle#1\right|}
\newcommand{\ket}[1]{\left|#1\right\rangle}
\newcommand{\bracket}[2]{\big\langle#1 \bigm| #2\big\rangle}
\newcommand{\Tr}{{\rm Tr}}
\renewcommand{\Im}{{\rm Im}}
\renewcommand{\Re}{{\rm Re}}
\newcommand{\ef}{{\epsilon_{\rm F}}}
\newcommand{\MC}[1]{\mathcal{#1}}
\newcommand{\pp}{{\prime\prime}}
\newcommand{\ppp}{{\prime\prime\prime}}
\newcommand{\pppp}{{\prime\prime\prime\prime}}

\newcommand{\tip}{{\rm T}}
\newcommand{\rest}{{\rm R}}
\newcommand{\gammat}{{\bm\gamma}_{\rm T}}
\newcommand{\Gammat}{{\bm\Gamma}_{\rm T}}
\newcommand{\Gammar}{{\bm\Gamma}_{\rm R}}

\author{David Jacob}
\email{david.jacob@ehu.eus}
\affiliation{Nano-Bio Spectroscopy Group and European Theoretical Spectroscopy Facility (ETSF), Departamento de F\'isica de Materiales, Universidad del Pa\'is Vasco UPV/EHU, Av. Tolosa 72, E-20018 San Sebasti\'an, Spain}
\affiliation{IKERBASQUE, Basque Foundation for Science, Mar\'ia D\'iaz de Haro 3, E-48013 Bilbao, Spain}

\author{Stefan Kurth} 
\email{stefan.kurth@ehu.eus}
\affiliation{Nano-Bio Spectroscopy Group and European Theoretical Spectroscopy Facility (ETSF), Departamento de F\'isica de Materiales, Universidad del Pa\'is Vasco UPV/EHU, Av. Tolosa 72, E-20018 San Sebasti\'an, Spain}
\affiliation{IKERBASQUE, Basque Foundation for Science, Mar\'ia D\'iaz de Haro 3, E-48013 Bilbao, Spain}
\affiliation{Donostia International Physics Center (DIPC), Paseo Manuel de Lardizabal 4, E-20018 San Sebasti\'an, Spain}

\title{Many-body spectral functions from steady state density functional theory}

\begin{abstract}
  We propose a scheme to extract the many-body spectral function of an
  interacting many-electron system from an equilibrium density functional
  theory (DFT) calculation. To this end we devise an ideal
  scanning tunneling microscope (STM) setup
  and employ the recently proposed steady-state DFT formalism (i-DFT)
  which allows to calculate the steady current through a nanoscopic
  region coupled to two biased electrodes. In our setup one of the electrodes
  serves as a probe ('STM tip'), which is weakly coupled to the system we
  want to measure. In the ideal STM limit of vanishing coupling to the tip, the
  system is restored to quasi-equilibrium and the normalized differential
  conductance yields the exact equilibrium many-body spectral function.
  Calculating this quantity from i-DFT, we derive an exact relation expressing
  the interacting spectral function in terms of the Kohn-Sham one. As
  illustrative examples we apply our scheme to calculate the spectral
  functions of two non-trivial model systems, namely the single Anderson impurity
  model and the Constant Interaction Model.
\end{abstract}

\maketitle

Density functional theory
(DFT)\cite{Hohenberg:PR:1964,Kohn:PR:1965,DreizlerGross:90} is without doubt one
of the most popular and succesful approches for the description of matter,
with important applications in condensed matter physics, material science and
computational chemistry. DFT owes its success to its relative simplicity and
low computational cost as compared to other approaches for solving the quantum
many-body problem. Despite its simplicity, DFT is in principle exact, i.e.,
it can provide the exact ground state energy and density of many-electron
systems. In practice, of course, an approximation for the exchange
correlation (xc) energy functional is needed and a plethora of such
approximations have been suggested
\cite{Kohn:PR:1965,PerdewWang:86,Becke:88-2,LeeYangParr:88,Becke:93-2,PerdewBurkeErnzerhof:96,PerdewKurthZupanBlaha:99,TaoPerdewStroverovScuseria:03,KuemmelKronik:08}.

Via the Hohenberg-Kohn theorem \cite{Hohenberg:PR:1964} the
ground state density uniquely determines the external potential (up to a
constant). Therefore the many-electron Hamiltonian and thus {\em all}
physical properties of the interacting system (including, e.g., excitation
energies or spectral functions) are determined in principle
uniquely by the ground state density. In practice, of course, this functional
dependence is unknown and these quantities have to be extracted from some
alternative theoretical framework. While optical excitations can successfully
be computed within time-dependent DFT (TDDFT)
\cite{RungeGross:84,Ullrich:12,Maitra:16},
spectral functions which encode information about the
(quasi-particle) excitations of the system (energies and lifetimes)
have so far been out of reach for DFT and instead are typically
calculated with a Green function framework
\cite{OnidaReiningRubio:02,MartinReiningCeperley:16}. 
Spectral functions to a large degree determine the 
transport properties of a many-electron system and can (approximately)
be measured, e.g., with STM spectroscopy or angular resolved
photoemisson spectroscopy (ARPES). 

Despite the lack of a formal justification, the eigenvalues 
of the fictitious
non-interacting Kohn-Sham (KS) system \cite{Kohn:PR:1965} of DFT are often
used as an approximation for the quasi-particle band structures of solids.
This approach works reasonably well for weakly correlated systems, but
fails for strongly correlated ones. An
extreme case is the one of the Mott-Hubbard insulator for which the KS
spectrum predicts a metallic ground state, even when using the exact xc
functional \cite{CapelleCampo:13}.

In the present work we describe a scheme to extract the {\em interacting}
spectral function (and thus an excited state property)
essentially from a ground
state DFT calculation. In order to do so, we need to make use of a recently
proposed DFT framework for non-equilibrium steady state transport, the
so-called i-DFT approach~\cite{StefanucciKurth:15,KurthStefanucci:17}. 
Under certain, well-defined conditions (see below) the i-DFT
self-consistent equations for density and current decouple. While the
one for the density becomes equivalent to the usual ground-state
DFT selfconsistency condition, the extra equation for the current can
be used to extract the spectral function.

\begin{figure}
  \begin{flushleft}
    \includegraphics[width=0.25\linewidth]{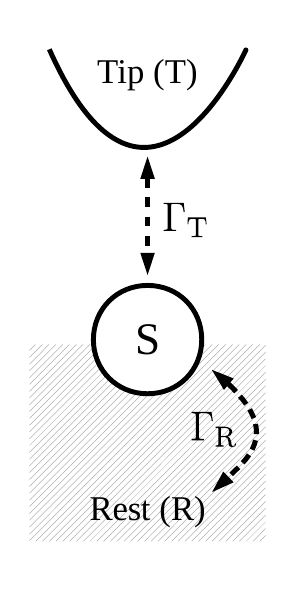}
    \includegraphics[width=0.72\linewidth]{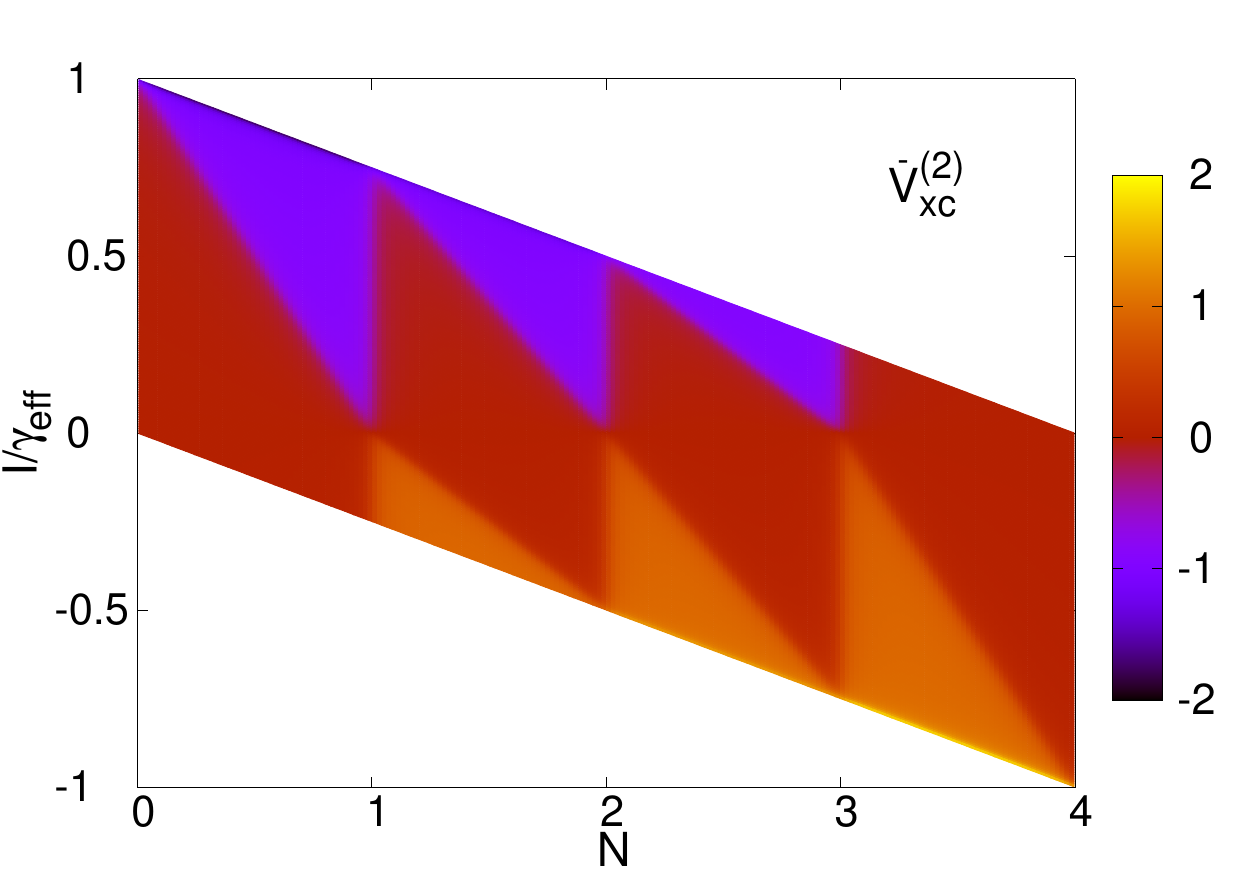}
  \end{flushleft}
  \caption{
    \label{fig:setup}
    Left: STM like theoretical setup for measuring the true many-body spectral
    function of a system in a density functional theory framework. The
    tip (T), couples very weakly to the sample (S), which in turn is strongly
    coupled to the rest of the system (R). Right: xc bias
    $\bar{V}_{\rm xc}^{(2)}$ of Eq.~(\ref{xcbias_CIM}) for the CIM with 
    $\mathcal{M}=2$ and $U/\gamma=5$.
  }
\end{figure}

The basic idea is to 'measure' the spectral function of a system by means of
an STM like setup where a small portion of the system ($S$) is probed by an
STM tip ($T$), as shown in the left panel of Fig.~\ref{fig:setup}. The tip couples only very
weakly to the sample $S$ and thus does not
influence the
system in an essential way. Hence the system $S$ to be probed is essentially
in equilibrium with electrode $R$. In addition, we assume that the
applied bias $V$ drops entirely at the STM tip. Then the Keldysh Green 
function (GF) \cite{svl-book} of the sample region becomes independent of the
bias.\footnote{The non-interacting retarded and advanced GFs
  $\bm{G}_0^{r,a}=(\omega-\bm{H}_0-\bm{\Sigma}^{r,a}_\tip-\bm{\Sigma}^{r,a}_\rest)
  ^{-1}$ are independent of the bias by construction, while the lesser GF is in
  principle bias-dependent via the tip Fermi function
  $f_\tip=f(\omega-V)$: $G_0^<=G_0^r(f_\tip\Gammat+f_\rest\Gammar)G_0^a$.
  However, in the limit $\Gammat\rightarrow0$ the bias dependence
  vanishes as $\partial{G}^<_0/\partial{V}\sim\Gammat$. As the interacting
  GFs can be written diagrammatically in terms of the non-interacting
  GFs and the electron-electron interaction does not depend on the bias
  either, it follows that the interacting GFs must also be 
  independent of the bias in the limit $\Gammat\rightarrow0$, i.e.
  $\partial{G}^{r,a,<}/\partial{V}\rightarrow0$.}
Thus the density matrix
$\bm\rho$ of $S$, and correspondingly the particle density
$n(\bm{r})=\sum_{m,m^\prime}\phi^\ast_m(\bm{r})\rho_{mm^\prime}\phi_{m^\prime}(\bm{r})$
(where the $\phi_m(\bm{r})$ form a single-electron basis spanning $S$) are
also independent of $V$, $\partial{\bm{\rho}}/\partial{V}=0$, and can be
calculated from the equilibrium expression,
\begin{equation}
  \bm{\rho} =  -2i\,\int\frac{d\omega}{2\pi} \bm{G}^<(\omega) \xrightarrow[\Gammat\to0]{} 2\,\int\frac{d\omega}{2\pi}\, f(\omega)\,\bm{A}(\omega)
\label{eq:dens_matrix}
\end{equation}
where $\bm{G}^<$ is the lesser GF matrix, $\bm{A}\equiv{i(\bm{G}^a-\bm{G}^r)}$ the spectral function,
$\bm{G}^a$ and $\bm{G}^r$ the retarded and advanced GF matrices, respectively, and $f(\omega)$ the
Fermi function of $S$+$R$ with chemical potential $\mu\equiv0$.

The current from the tip to the sample is given by the Meir-Wingreen expression \cite{Meir:PRL:1992}:
\begin{equation}
  \label{eq:Meir-Wingreen}
  I(V) = 2\, \int\frac{d\omega}{2\pi}\, \Tr\left\{ f_\tip(\omega)\,\Gammat\,\bm{A}(\omega) + i \Gammat\, \bm{G}^<(\omega) \right\}
\end{equation}
where $f_\tip(\omega)=f(\omega-V)$
and we have defined the coupling matrix
$\bm{\Gamma}_\alpha=i(\bm{\Sigma}_\alpha^\dagger-\bm{\Sigma}_\alpha)$ of lead
$\alpha$ ($\alpha={\rm T,R}$ for tip and rest, respectively) expressed in
terms of the corresponding embedding self energy $\bm{\Sigma}_\alpha$. 

By taking the derivative of Eq.~(\ref{eq:Meir-Wingreen}) w.r.t. $V$ in the
ideal STM limit of vanishing coupling to the tip, $\Gammat\rightarrow0$, and
with $V$ dropping entirely at the tip, we find that the differential
conductance can be expressed solely in terms of the \emph{equilibrium}
spectral function of the sample:
\begin{equation}
  \label{eq:dIdV}
  \frac{\partial{I}}{\partial{V}} \xrightarrow[\Gammat\to0]{} \int\frac{d\omega}{\pi} \,
  \frac{\partial{f_\tip}}{\partial{V}} \,\Tr\left[ \Gammat \bm{A}(\omega) \right]
  \xrightarrow[T\to0]{} \frac{\Tr\left[ \Gammat \bm{A}(V) \right]}{\pi}
  %(-f^\prime(\omega-V)) 
\end{equation}
where in the last step we have used that in the zero-temperature limit 
$T\rightarrow0$ the derivative of the Fermi function becomes a 
$\delta$-function,
$\partial{f_\tip}/\partial{V}=-f^\prime(\omega-V)\rightarrow\delta(\omega-V)$. 
Eq.~(\ref{eq:dIdV}) holds for arbitrary coupling $\Gammat$ to the tip. 
By choosing the coupling matrix such that only a single matrix element is 
nonvanishing, i.e., $\Gammat=\gamma^{{\rm T}}_{lm}\ket{l}\bra{m}$, one can thus 
extract an arbitrary matrix element of the (many-body) spectral function 
matrix as
\begin{equation}
  \label{eq:specfunc}
  A_{ml}(\omega) = \lim_{\gamma^{{\rm T}}_{lm} \to 0} \frac{\pi}{\gamma^{{\rm T}}_{lm}} 
  \left.\frac{\partial{I}}{\partial{V}}\right|_{V=\omega} 
\end{equation}
provided that the $I-V$ characteristic of the interacting system 
is known.

Here we choose a recently proposed density functional theory 
for steady-state transport \cite{StefanucciKurth:15}, named i-DFT, as 
the framework for calculating the steady current $I$ and then exploit 
Eq.~(\ref{eq:specfunc}) to extract the spectral function. The central idea of 
i-DFT is that the pair of ``densities'' $(n(\bm{r}),I)$ of the interacting 
system can in principle exactly be reproduced by an effective 
system of non-interacting electrons, the KS system. This KS system features 
both a local Hartree-exchange-correlation (Hxc) potential 
$v_{\rm Hxc}[n,I](\bm{r})$ in $S$ as well as a (spatially constant) 
exchange-correlation (xc) contribution $V_{\rm xc}[n,I]$ to the bias. Both 
these (H)xc potentials are functionals of the density $n(\bm{r})$ in $S$ 
and the steady current $I$. 
 
Originally, the self-consistent KS equations of i-DFT for the steady state 
density and current were formulated for the situation of a bias applied 
symmetrically in both leads \cite{StefanucciKurth:15} but they are easily 
transformed to the situation where the applied bias $V$ drops entirely at 
the tip.\footnote{Transformation from symmetric voltage drop $V_\tip=-V_\rest=V/2$ 
  to a completely asymmetric voltage drop, $V_\tip=V$ and $V_\rest=0$, is 
  achieved by a spatially constant shift of the gate potential by $-V/2$, and 
  subsequent substitution of the integration variable, 
  $\omega\rightarrow\omega+(V+V_{\rm xc})/2$.}
For this asymmetrically applied bias and arbitrary couplings $\Gammat$ and 
$\Gammar$ to the electrodes, the self-consistent i-DFT KS equations read 
\begin{subequations}
\begin{eqnarray}
  n(\bm{r}) &=& 2\int\frac{d\omega}{2\pi} \left[ f(\omega-V_s) 
    A_{\tip,s}\left(\bm{r};\omega\right) + f(\omega) 
    A_{\rest,s}\left(\bm{r};\omega\right) \right] 
  \label{eq:dens}\nonumber\\
  \\
  I &=& 2\, \int\frac{d\omega}{2\pi}\, [f(\omega-V_s)-f(\omega)] \, 
  T_s\left(\omega\right) \hspace{1ex}
  \label{eq:curr}
\end{eqnarray}
\end{subequations}
where $\bm{A}_{\alpha,s}=\bm{G}_s\bm{\Gamma}_\alpha\bm{G}_s^{\dagger}$ is the 
non-equilibrium KS spectral function of $S$ associated with electron injection 
from electrode $\alpha$, 
$A_{\alpha,s}(\bm{r};\omega)=\bra{\bm{r}}\bm{A}_{\alpha,s}(\omega)\ket{\bm{r}}$
its spatial representation. $T_s=\Tr[\Gammat\bm{G}_s^\dagger\Gammar\bm{G}_s]$ 
is the KS transmission function and
$\bm{G}_s=(\omega-\frac{V_{\rm xc}}{2} -\bm{h}_s - 
\bm\Sigma_\tip-\bm\Sigma_\rest)^{-1}$
is the (retarded) non-equilibrium KS Green 
function of the sample region. Here $\bm{h}_s = \bm{t} + \bm{v}_s$ is the KS 
Hamiltonian in matrix notation with $\bm{t}$ the kinetic energy and 
$\bm{v}_s$ the KS ``gate'' potential which, in the position basis 
is given as usual by $v_s(\bm{r})=v(\bm{r}) + v_{\rm Hxc}(\bm{r})$. 
$V_s = V + V_{\rm xc}$ is the effective KS bias containing both 
the externally applied bias $V$ and the xc bias $V_{\rm xc}$.
Note that here the frequency dependence of the embedding
self energies and the corresponding coupling matrices is given by
$\bm\Sigma_\tip = \bm\Sigma_\tip(\omega - V_s)$ for the tip and
$\bm\Sigma_\rest = \bm\Sigma_\rest(\omega)$ for the rest where
$\bm\Sigma_{\alpha}(\omega)$ is the embedding energy of lead $\alpha$ in
equilibrium.

In the ideal STM limit, $\Gammat\rightarrow0$, the i-DFT expression 
(\ref{eq:dens}) for the density reduces to $n(\bm{r}) = 
2\,\int\frac{d\omega}{2\pi}\, f(\omega) A_{{\rm R},s}(\bm{r};\omega)$. As we 
have seen in Eq.~(\ref{eq:dens_matrix}), in this limit the density matrix and 
thus the density take on their equilibrium values and become independent of 
the applied bias, $\partial{n}(\bm{r})/\partial{V}=0$.
The equilibrium density can be expressed in terms of the
{\em equilibrium} KS spectral function $A_s^{(0)}(\omega)$ as
\begin{equation}
  \label{dens}
  n(\bm{r}) = \int\frac{d\omega}{\pi}\, f(\omega) A_s^{(0)}(\bm{r};\omega) \;.
\end{equation}
Since the equilibrium 
Hxc potential $v_{\rm Hxc}^{(0)}[n](\bm{r})$ which yields the exact equilibrium 
density is unique, we can thus deduce the following important relationship 
between the xc bias and Hxc gate:
\begin{equation}
  \label{eq:condition}
  \lim_{\Gammat\rightarrow0} v_{\rm Hxc}[n,I](\bm{r}) + 
  \frac{1}{2}V_{\rm xc}[n,I] = v_{\rm Hxc}^{(0)}[n](\bm{r}) \;.
\end{equation}
As a consequence  of the ideal STM limit, 
the i-DFT equation for the density is completely decoupled from the 
current and can be solved at equilibrium, i.e. within a normal ground state 
KS DFT calculation. Furthermore, for $\Gammat\rightarrow0$ the transmission 
function reduces to an expression similar to Eq.~(\ref{eq:dIdV}), i.e., 
$T_s\rightarrow\Tr[\Gammat\bm{A}_s^{(0)}]$, where $\bm{A}_s^{(0)}(\omega)$ 
is the \emph{equilibrium} KS spectral function. For the 
tip coupling matrix we again take $\Gammat=\gamma^{{\rm T}}_{lm}\ket{l}\bra{m}$ 
and, using Eq.~(\ref{eq:condition}), obtain 
\begin{equation}
  \label{eq:current}
  \lim_{\gamma^{\rm T}_{lm}\to 0} \frac{I}{\gamma^{\rm T}_{lm}} = 
\int\frac{d\omega}{\pi}\, [f(\omega-V-V_{\rm xc})-f(\omega)] 
  \, A_{ml,s}^{(0)}(\omega) \;.
\end{equation}
Note that the dependence on the bias shows up explicitly in the Fermi function 
of the tip and implicitly in the xc bias $V_{\rm xc}$ which depends on the 
current. In other words, for a given external gate potential (and thus at fixed 
equilibrium density $n(\bm{r})$ and KS spectral function 
$\bm{A}_s^{(0)}(\omega)$), Eq.~(\ref{eq:current}) becomes 
a self-consistency condition for the current. From Eq.~(\ref{eq:current}) we 
can calculate the differential conductance and, via Eq.~(\ref{eq:specfunc}), 
determine the many-body spectral function in the limit of zero temperature. 
Taking the derivative of Eq.~(\ref{eq:current}) w.r.t. $V$ in the zero
temperature limit yields
\begin{equation}
  A_{ml}(\omega) = \lim_{\gamma^{\rm T}_{lm}\to 0} 
  \left( 1 + \frac{\gamma_{lm}^{\rm T}}{\pi} 
  \frac{\partial{V_{\rm xc}}}{\partial{I}} A_{ml}(\omega)
  \right) A_{ml,s}^{(0)}(\omega+V_{\rm xc}) \;.
\end{equation}
where we have taken into account that $V_{\rm xc}$ depends on $I$.
Solving for $A_{ml}(\omega)$ we arrive at the central result of our paper 
which relates the many-body spectral function to the equilibrium KS spectral 
function:
\begin{equation}
  \label{eq:relation}
  A_{ml}(\omega) = \lim_{\gamma^{\rm T}_{lm}\to 0} 
  \frac{ A_{ml,s}^{(0)}(\omega+V_{\rm xc}) }{1-
    \frac{\gamma_{lm}^{\rm T}}{\pi}\frac{\partial{V_{\rm xc}}}{\partial{I}} 
    A_{ml,s}^{(0)}(\omega+V_{\rm xc}) }
\end{equation}
The xc bias $V_{\rm xc}$ and its derivative $\partial{V_{\rm xc}}/\partial{I}$ are to 
be evaluated with the current $I$ obtained by solving Eq.~(\ref{eq:current}) at 
bias $V=\omega$. The appearance of $\gamma^{\rm T}_{lm}$ as prefactor of the
second term in the denominator of Eq.~(\ref{eq:relation})  
does {\em not} imply that this term vanishes in the limit 
$\gamma^{\rm T}_{lm} \to 0$ since $\frac{\partial{V_{\rm xc}}}{\partial{I}}$
diverges as $1/\gamma^{\rm T}_{lm}$ (see example for $V_{\rm xc}$ below).
We emphasize that Eq.~(\ref{eq:relation}) is an exact result provided that 
the exact functionals $v_{\rm Hxc}[n]$ and $V_{\rm xc}[n,I]$ are used. In 
practice, of course, these functionals are unknown and need to be 
approximated (see below).
The proposed scheme to calculate the 
spectral function is computationally very efficient: it requires only the
usual KS self-consistency for the density plus, for any frequency (bias),
the solution of Eq.~(\ref{eq:current}) for the current. 
Eq.~(\ref{eq:relation}) expresses the equilibrium 
spectral function as functional of the ground state density using concepts 
of i-DFT (the xc bias $V_{\rm xc}$). This is similar to TDDFT in the linear regime
where the linear density response function is  
a functional of the ground state density expressed in terms of the 
(frequency-dependent) xc kernel, a pure TDDFT quantity.

In the following we apply our formalism to model systems $S$ to be probed by 
the STM setup. We model $S$ as a quantum dot described by the 
constant interaction model (CIM) for which the Hamiltonian is 
$\hat{H}^{\rm CIM} = \sum_{i \sigma} 
\varepsilon_i \hat{n}_{i\sigma} + U/2 \sum_{i \sigma\neq j \sigma'} \hat{n}_{i\sigma} 
\hat{n}_{j\sigma'}$ where $\hat{n}_{i\sigma}$ is the electron occupation operator 
for level $i$ with spin $\sigma$. The system $S$ is connected to both a tip 
$T$ and a second lead $R$ via energy independent couplings $\Gamma_{\alpha}$, 
i.e., we are in the wide-band limit (WBL) for both leads. Note that for a 
single level this becomes the single-impurity Anderson model (SIAM) 
\cite{Anderson:61}. The CIM has been studied within the i-DFT framework both 
in the Coulomb blockade \cite{StefanucciKurth:15} as well as in the Kondo 
regime \cite{KurthStefanucci:16,KurthStefanucci:17} and approximate i-DFT 
xc potentials have been suggested.
For simplicity, we restrict ourselves to the CIM with an arbitrary number
$\mathcal{M}$ of degenerate single-particle levels
($\varepsilon_i=\varepsilon$) which are 
all coupled in the same way to the lead $\alpha$ , i.e., the coupling 
matrices in the single-particle 
basis $\Gamma_{\alpha} = \gamma^{\alpha} {\mathbf 1}$ are proportional to the 
unit matrix $\mathbf 1$ (the constants $\gamma^{\rm T}$ and $\gamma^{\rm R}$ can 
differ). In this case the i-DFT xc potentials depend only on the total number 
$N=\sum_{i\sigma} n_{i\sigma}$ of electrons on the dot.

We need to solve the usual DFT self-consistency for the ground state density
and, as usual, in order to do so we need an approximation for the Hxc potential
$v^{(\mathcal{M})}_{\rm Hxc}[N]$ of the degenerate ${\mathcal{M}}$-level CIM. The
general structure of this potential is clear: $v^{(\mathcal{M})}_{\rm Hxc}[N]$
will exhibit steps of height $U$ at integer $N$. In the language of DFT,
the height $U$ of these steps can be identified with the famous derivative
discontinuity of the xc energy functional \cite{PerdewParrLevyBalduz:82}.
For the uncontacted CIM at zero temperature, the exact Hxc potential is
discontinuous \cite{StefanucciKurth:13} but when the CIM is brought in
contact with (wide band) leads, these discontinuities are smoothened, the
smoothening governed by the parameter $\gamma=\gamma^{\rm R}$. In the present
work we use for $v^{(\mathcal{M})}_{\rm Hxc}[N]$ the form suggested in
Eqs.~(143) and (144) of Ref.~\cite{KurthStefanucci:17} which consists
of a sum of accurately parametrized Hxc potentials for the SIAM
\cite{blbs.2012}. 

We still need an approximate functional for the xc bias
$V^{(\mathcal{M})}_{\rm xc}[N,I]$ of the degenerate ${\mathcal{M}}$-level CIM in
the limit $\gamma^{\rm T} \to 0$. 
Below we propose an approximation and discuss the various ideas 
entering in its construction. Our approximate functional reads 
\begin{equation}
V^{(\mathcal{M})}_{\rm xc}[N,I] = (1 - a[I]) \bar{V}^{(\mathcal{M})}_{\rm xc}[N,I] \;.
\label{eq:xcbias_CIM_mod}
\end{equation}
where $a[I]$ is a purely current-dependent function for which we choose
the form
\begin{equation}
  a[I] = 1 - \frac{2}{\pi} \arctan\left[
    \lambda \left(\frac{I}{W  \gamma_{\rm eff}} \right)^2 \right]
  \label{eq:afunc}
\end{equation}
with the parameters $\lambda=0.16$, $\gamma_{\rm eff}=\frac{4\gamma^{\rm T}\gamma^{\rm R}}{\gamma^{\rm T}+\gamma^{\rm R}}$
and $\bar{V}^{(\mathcal{M})}_{\rm xc}[N,I]$
given as 
\begin{equation}
  \bar{V}^{(\mathcal{M})}_{\rm xc}[N,I] = -\sum_{K=1}^{2\mathcal{M}-1} U
  \sum_{s=\pm}\frac{s}{\pi}\,
      {\rm atan}\left(\frac{\Delta_{K}^{(s)}(N,I)}{2 W}\right) 
      \label{xcbias_CIM}
\end{equation}
with $W=0.16 \gamma/U$. 
The $\Delta_{K}^{(s)}(N,I)$, $s=\pm$, are piecewise linear functions of $N$ and
$I$. For $s=+$ they are independent of the current and read
\begin{equation}
  \Delta_{K}^{(+)}(N,I) = N - K
\end{equation}
while for $s=-$ the corresponding form is 
\begin{equation}
  \Delta_{K}^{(-)}(N,I) = \left\{
  \begin{array}{cl}
    N + \alpha_{K}^{(+)} \frac{I}{4 \gamma^{\rm T}} - K & \mbox{for $I \geq 0$} \\
    N + \alpha_{K}^{(-)} \frac{I}{4 \gamma^{\rm T}} - K & \mbox{for $I <0$}
  \end{array} \right.
\end{equation}
The constants $\alpha_{K}^{(s)}$, $s=\pm$, are given by
\begin{equation} 
    \alpha_{K}^{(+)} = \frac{4}{2\mathcal{M}-K+1}
    \hspace{1ex}\mbox{ and }\hspace{1ex}
    \alpha_{K}^{(-)} = \frac{4}{(K+1)} \;.
  \end{equation}
For completeness we note the convention used throughout that a positive
current flows from the tip $T$ to the sample $S$.

Where do the different ingredients for the xc bias come from?
In Ref.~\cite{StefanucciKurth:15} we constructed xc functionals for the 
Coulomb blockade regime for a symmetrically coupled, degenerate
$\mathcal{M}$-level CIM by numerical
inversion of rate equations \cite{rate-paper1}. The resulting xc potentials
showed a complex pattern of smeared steps of height $U/2$ for the Hxc gate
and height $U$ for the xc bias potential. The position of the steps is
determined by piecewise linear functions connecting vertices in the $N-I$
plane where the vertices correspond to $(2\mathcal{M}+1)^2$ plateau values of
density and current in a scan over gate and bias. Here we are interested in
the situation of a CIM with completely asymmetric coupling
$\gamma^{\rm T} \to 0$. By analyzing the rate equations in this case we found
that (i) the codomain of the Hxc potentials becomes a parallelogram connecting
the points $(N,I/\gamma_{\rm eff})=(0,0), (0,\mathcal{M}/2), (2\mathcal{M},0),
(2 \mathcal{M},-\mathcal{M}/2)$ and (ii) all the vertices except for the ones
with vanishing current (and integer density) are pushed to the boundaries of
the domain. As a consequence, the resulting Hxc potentials are
significantly less complex than for symmetric coupling.
As an example, in the right panel of
Fig.~\ref{fig:setup} we show the Hxc bias 
$\bar{V}^{(2)}_{\rm xc}[N,I]$ for an asymmetrically coupled two-level CIM.
The crucial feature in both Hxc gate and xc bias are
the smeared steps which are directly related to the derivative
discontinuity of DFT.

The model xc bias $\bar{V}^{(\mathcal{M})}_{\rm xc}[N,I]$ contains Coulomb blockade
but no Kondo physics. In a DFT framework, the Kondo effect in the zero-bias 
conductance of weakly coupled quantum dots is already captured correctly in 
the KS conductance, both for single-level 
\cite{sk.2011,blbs.2012,tse.2012} as well as for multi-level dots 
\cite{StefanucciKurth:13}. The incorporation of Kondo physics into an i-DFT 
functional thus requires that the derivative of the xc bias w.r.t. the current 
vanishes for $I=0$ \cite{StefanucciKurth:15} which can be achieved with the
ansatz of Eq.~(\ref{eq:xcbias_CIM_mod}) if $a[I=0]=1$. Here we choose the
functional form of $a[I]$ as given in Eq.~(\ref{eq:afunc}) which is slightly
different from the one used 
in Refs.~\cite{KurthStefanucci:16,KurthStefanucci:17}.

\begin{figure}
  \includegraphics[width=\linewidth]{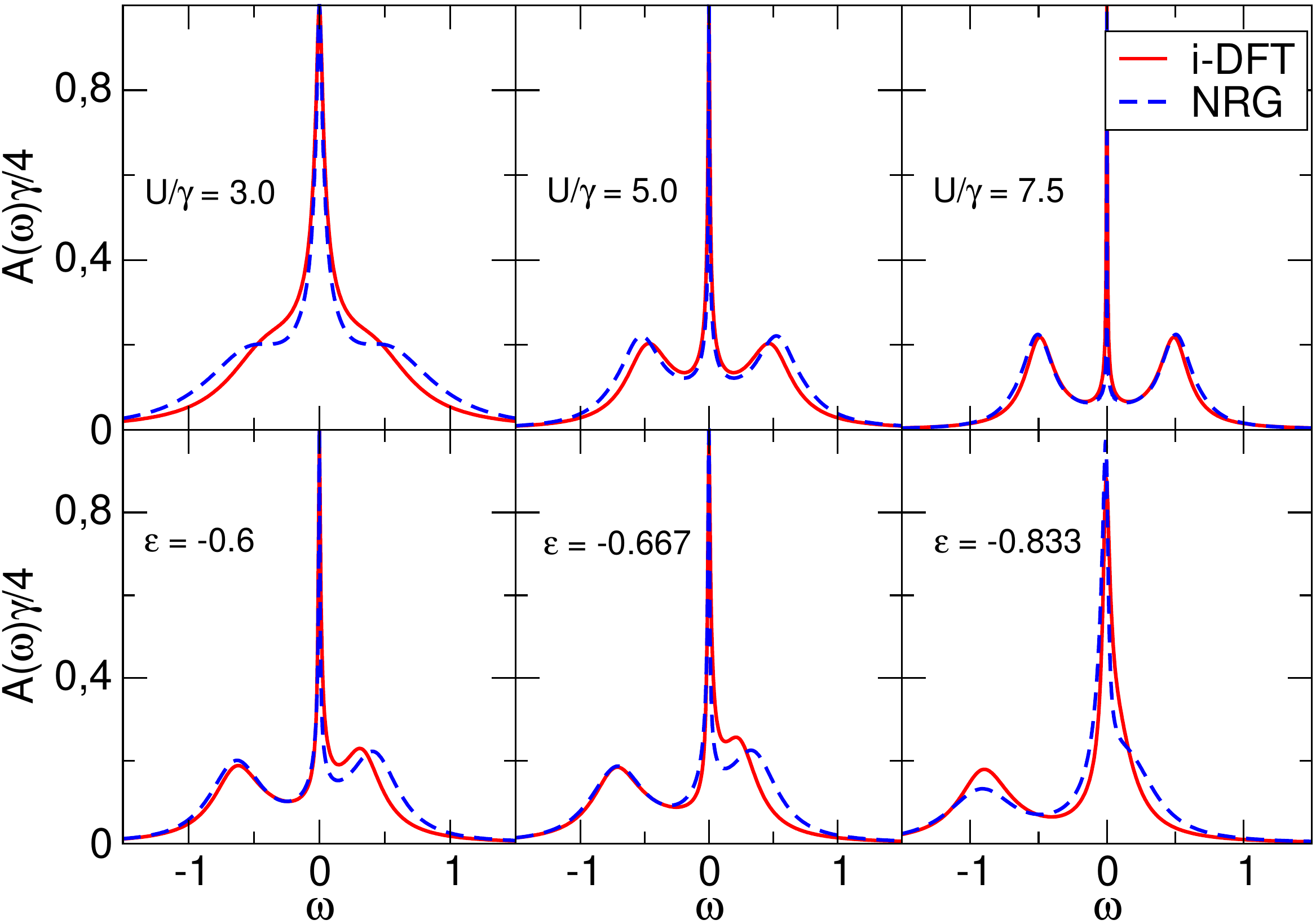}
  \caption{
    \label{fig:SIAM}
    Comparison of Anderson model spectra calculated from i-DFT via
    Eq. (\ref{eq:relation}) with NRG (taken from
    Ref.~\cite{Motahari:PRB:2016}). Upper panels: at the particle-hole
    symmetric point for different values of $U/\gamma$. Lower panels: for
    $U/\gamma=5$ at different values $\varepsilon$ for the on-site energy.
    Energies are given in units of $U$.
  }
\end{figure}

\begin{figure}
  \includegraphics[width=0.99\linewidth]{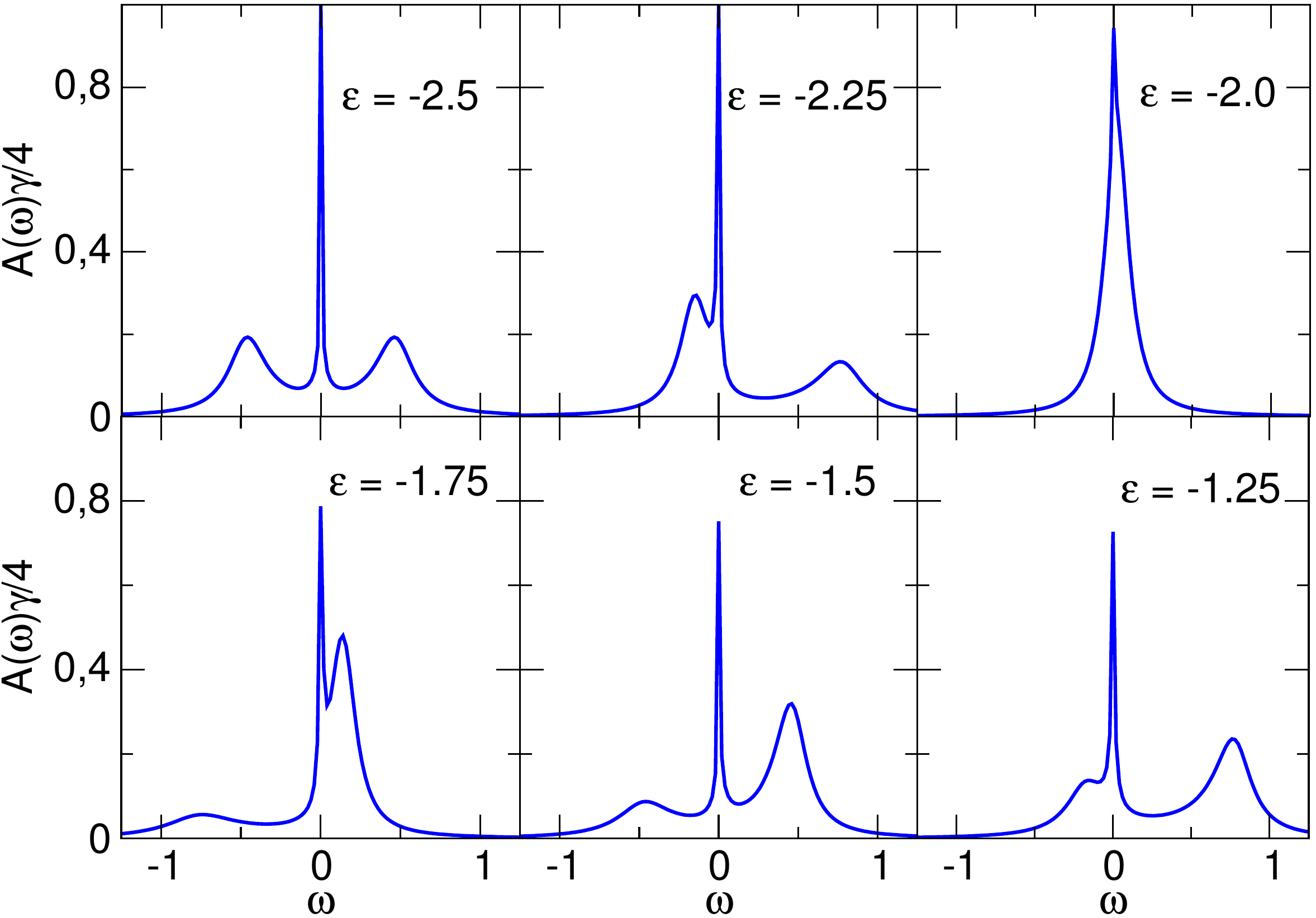}
  \caption{\label{fig:CIM}
    Spectral functions obtained from i-DFT for the CIM with three
    degenerate single-particle levels for $U/\gamma=8$ and different values
    of the gate potential. Energies in units of $U$.}
\end{figure}

As an illustration of the quality of the obtained i-DFT results, we calculate
spectral functions for the SIAM both at and away from particle-hole symmetry
(Fig.~\ref{fig:SIAM}) and compare to numerically accurate
Numerical Renormalization Group (NRG) 
results \cite{Bulla:RMP:2008,Motahari:PRB:2016}.
We see that our i-DFT functional captures the
essential spectral features in all cases although sometimes the height and
position of the side peaks is slightly off, especially for weak interactions.
For strong interactions (upper right panel) the agreement of the i-DFT
spectrum with the NRG one is quite remarkable.
In Fig.~\ref{fig:CIM} we show
the i-DFT spectral functions for a degenerate 3-level CIM at various values
of the gate. In most cases, the spectra are qualitatively similar to SIAM
spectra. Only for $\varepsilon=-2$ (upper right panel) there is just a single
spectral peak which essentially comes from one of the side peaks merging
with the Kondo resonance as the position of the side peak changes from below
to above the Fermi level as the gate is increased. We emphasize
that the deviations of the i-DFT spectra from the exact ones have to be
attributed {\em entirely} to the approximations we used for the Hxc gate and
xc bias potentials. If their exact forms were used, the
i-DFT spectra would be exact.
We also note that the xc bias of Eq.~(\ref{xcbias_CIM}) lends
itself to straightforward generalization beyond the CIM by replacing
$U$ with $U(K)\equiv\Delta_{\rm xc}(K)$ where we identified
the charging energy $U(K)$ of charging state $K$ with the
derivative discontinuity $\Delta_{\rm xc}(K)$ of the isolated dot with $K$
electrons. In this way it may be possible to construct an xc bias functional
from a standard density functional provided the latter yields a
non-vanishing derivative discontinuity
\cite{KraislerKronik:13,BaerNeuhauser:05,YamadaFengHoskinsFenkDunietz:16}.

In summary, we have proposed a computationally efficient
DFT scheme to calculate the spectral function
of an {\em interacting} many-electron system. Conceptually, this scheme
allows to express the spectral function in terms of the ground state density
(completely in line with the Hohenberg-Kohn theorem) although we use
concepts from a DFT formulation for steady state transport (i-DFT).
From the many-body spectral function one can obtain the many-body self-energy
and the Green function which in our scheme therefore also allows to express
them as functionals of the ground state density. In the derivation of our scheme
and also in the applications shown here, we considered the typical
transport setup of a system connected to two leads, one of them being the
weakly coupled tip. However, the scheme should also be applicable to the
calculation of surface and even bulk spectral functions by considering
the system $S$ and the rest $R$ together while the tip just becomes a
computational device to extract the spectral function. This idea as well
as the corresponding construction of approximate functionals are subject
of ongoing work.

\begin{acknowledgments}
We would like to thank Gianluca Stefanucci for useful discussions.
We acknowledge funding through the grant
``Grupos Consolidados UPV/EHU del Gobierno Vasco'' (IT578-13). 
S.K. additionally acknowledges funding through a grant of the
"Ministerio de Economia y Competividad (MINECO)" (FIS2016-79464-P). 
\end{acknowledgments}

\bibliography{nanodmft,nano_bibfile}

\begin{thebibliography}{36}%
\makeatletter
\providecommand \@ifxundefined [1]{%
 \@ifx{#1\undefined}
}%
\providecommand \@ifnum [1]{%
 \ifnum #1\expandafter \@firstoftwo
 \else \expandafter \@secondoftwo
 \fi
}%
\providecommand \@ifx [1]{%
 \ifx #1\expandafter \@firstoftwo
 \else \expandafter \@secondoftwo
 \fi
}%
\providecommand \natexlab [1]{#1}%
\providecommand \enquote  [1]{``#1''}%
\providecommand \bibnamefont  [1]{#1}%
\providecommand \bibfnamefont [1]{#1}%
\providecommand \citenamefont [1]{#1}%
\providecommand \href@noop [0]{\@secondoftwo}%
\providecommand \href [0]{\begingroup \@sanitize@url \@href}%
\providecommand \@href[1]{\@@startlink{#1}\@@href}%
\providecommand \@@href[1]{\endgroup#1\@@endlink}%
\providecommand \@sanitize@url [0]{\catcode `\\12\catcode `\$12\catcode
  `\&12\catcode `\#12\catcode `\^12\catcode `\_12\catcode `\%12\relax}%
\providecommand \@@startlink[1]{}%
\providecommand \@@endlink[0]{}%
\providecommand \url  [0]{\begingroup\@sanitize@url \@url }%
\providecommand \@url [1]{\endgroup\@href {#1}{\urlprefix }}%
\providecommand \urlprefix  [0]{URL }%
\providecommand \Eprint [0]{\href }%
\providecommand \doibase [0]{http://dx.doi.org/}%
\providecommand \selectlanguage [0]{\@gobble}%
\providecommand \bibinfo  [0]{\@secondoftwo}%
\providecommand \bibfield  [0]{\@secondoftwo}%
\providecommand \translation [1]{[#1]}%
\providecommand \BibitemOpen [0]{}%
\providecommand \bibitemStop [0]{}%
\providecommand \bibitemNoStop [0]{.\EOS\space}%
\providecommand \EOS [0]{\spacefactor3000\relax}%
\providecommand \BibitemShut  [1]{\csname bibitem#1\endcsname}%
\let\auto@bib@innerbib\@empty
%</preamble>
\bibitem [{\citenamefont {Hohenberg}\ and\ \citenamefont
  {Kohn}(1964)}]{Hohenberg:PR:1964}%
  \BibitemOpen
  \bibfield  {author} {\bibinfo {author} {\bibfnamefont {P.}~\bibnamefont
  {Hohenberg}}\ and\ \bibinfo {author} {\bibfnamefont {W.}~\bibnamefont
  {Kohn}},\ }\href {\doibase 10.1103/PhysRev.136.B864} {\bibfield  {journal}
  {\bibinfo  {journal} {Phys. Rev.}\ }\textbf {\bibinfo {volume} {136}},\
  \bibinfo {pages} {B864} (\bibinfo {year} {1964})}\BibitemShut {NoStop}%
\bibitem [{\citenamefont {Kohn}\ and\ \citenamefont
  {Sham}(1965)}]{Kohn:PR:1965}%
  \BibitemOpen
  \bibfield  {author} {\bibinfo {author} {\bibfnamefont {W.}~\bibnamefont
  {Kohn}}\ and\ \bibinfo {author} {\bibfnamefont {L.~J.}\ \bibnamefont
  {Sham}},\ }\href {\doibase 10.1103/PhysRev.140.A1133} {\bibfield  {journal}
  {\bibinfo  {journal} {Phys. Rev.}\ }\textbf {\bibinfo {volume} {140}},\
  \bibinfo {pages} {A1133} (\bibinfo {year} {1965})}\BibitemShut {NoStop}%
\bibitem [{\citenamefont {Dreizler}\ and\ \citenamefont
  {Gross}(1990)}]{DreizlerGross:90}%
  \BibitemOpen
  \bibfield  {author} {\bibinfo {author} {\bibfnamefont {R.~M.}\ \bibnamefont
  {Dreizler}}\ and\ \bibinfo {author} {\bibfnamefont {E.~K.~U.}\ \bibnamefont
  {Gross}},\ }\href@noop {} {\emph {\bibinfo {title} {Density Functional
  Theory}}}\ (\bibinfo  {publisher} {Springer},\ \bibinfo {address} {Berlin},\
  \bibinfo {year} {1990})\BibitemShut {NoStop}%
\bibitem [{\citenamefont {{J.P.~Perdew}}\ and\ \citenamefont
  {Wang}(1986)}]{PerdewWang:86}%
  \BibitemOpen
  \bibfield  {author} {\bibinfo {author} {\bibnamefont {{J.P.~Perdew}}}\ and\
  \bibinfo {author} {\bibfnamefont {Y.}~\bibnamefont {Wang}},\ }\href {\doibase
  10.1103/PhysRevB.33.8800} {\bibfield  {journal} {\bibinfo  {journal} {Phys.
  Rev. B}\ }\textbf {\bibinfo {volume} {33}},\ \bibinfo {pages} {8800}
  (\bibinfo {year} {1986})},\ \bibinfo {note} {ibid. {\bf 40}, 3399 (1989)
  (E)}\BibitemShut {NoStop}%
\bibitem [{\citenamefont {{A.D.~Becke}}(1988)}]{Becke:88-2}%
  \BibitemOpen
  \bibfield  {author} {\bibinfo {author} {\bibnamefont {{A.D.~Becke}}},\ }\href
  {\doibase 10.1063/1.454274} {\bibfield  {journal} {\bibinfo  {journal} {J.
  Chem. Phys.}\ }\textbf {\bibinfo {volume} {88}},\ \bibinfo {pages} {1053}
  (\bibinfo {year} {1988})}\BibitemShut {NoStop}%
\bibitem [{\citenamefont {Lee}\ \emph {et~al.}(1988)\citenamefont {Lee},
  \citenamefont {Yang},\ and\ \citenamefont {{R.G.~Parr}}}]{LeeYangParr:88}%
  \BibitemOpen
  \bibfield  {author} {\bibinfo {author} {\bibfnamefont {C.}~\bibnamefont
  {Lee}}, \bibinfo {author} {\bibfnamefont {W.}~\bibnamefont {Yang}}, \ and\
  \bibinfo {author} {\bibnamefont {{R.G.~Parr}}},\ }\href {\doibase
  10.1103/PhysRevB.37.785} {\bibfield  {journal} {\bibinfo  {journal} {Phys.
  Rev. B}\ }\textbf {\bibinfo {volume} {37}},\ \bibinfo {pages} {785} (\bibinfo
  {year} {1988})}\BibitemShut {NoStop}%
\bibitem [{\citenamefont {{A.D.~Becke}}(1993)}]{Becke:93-2}%
  \BibitemOpen
  \bibfield  {author} {\bibinfo {author} {\bibnamefont {{A.D.~Becke}}},\ }\href
  {\doibase 10.1063/1.464913} {\bibfield  {journal} {\bibinfo  {journal} {J.
  Chem. Phys.}\ }\textbf {\bibinfo {volume} {98}},\ \bibinfo {pages} {5648}
  (\bibinfo {year} {1993})}\BibitemShut {NoStop}%
\bibitem [{\citenamefont {{J.P.~Perdew}}\ \emph {et~al.}(1996)\citenamefont
  {{J.P.~Perdew}}, \citenamefont {Burke},\ and\ \citenamefont
  {Ernzerhof}}]{PerdewBurkeErnzerhof:96}%
  \BibitemOpen
  \bibfield  {author} {\bibinfo {author} {\bibnamefont {{J.P.~Perdew}}},
  \bibinfo {author} {\bibfnamefont {K.}~\bibnamefont {Burke}}, \ and\ \bibinfo
  {author} {\bibfnamefont {M.}~\bibnamefont {Ernzerhof}},\ }\href {\doibase
  10.1103/PhysRevLett.77.3865} {\bibfield  {journal} {\bibinfo  {journal}
  {Phys. Rev. Lett.}\ }\textbf {\bibinfo {volume} {77}},\ \bibinfo {pages}
  {3865} (\bibinfo {year} {1996})},\ \bibinfo {note} {ibid. {\bf 78}, 1396
  (1997)(E)}\BibitemShut {NoStop}%
\bibitem [{\citenamefont {{J.P.~Perdew}}\ \emph {et~al.}(1999)\citenamefont
  {{J.P.~Perdew}}, \citenamefont {Kurth}, \citenamefont {Zupan},\ and\
  \citenamefont {Blaha}}]{PerdewKurthZupanBlaha:99}%
  \BibitemOpen
  \bibfield  {author} {\bibinfo {author} {\bibnamefont {{J.P.~Perdew}}},
  \bibinfo {author} {\bibfnamefont {S.}~\bibnamefont {Kurth}}, \bibinfo
  {author} {\bibfnamefont {A.}~\bibnamefont {Zupan}}, \ and\ \bibinfo {author}
  {\bibfnamefont {P.}~\bibnamefont {Blaha}},\ }\href {\doibase
  10.1103/PhysRevLett.82.2544} {\bibfield  {journal} {\bibinfo  {journal}
  {Phys. Rev. Lett.}\ }\textbf {\bibinfo {volume} {82}},\ \bibinfo {pages}
  {2544} (\bibinfo {year} {1999})},\ \bibinfo {note} {ibid. {\bf 82}, 5179
  (1999)(E)}\BibitemShut {NoStop}%
\bibitem [{\citenamefont {Tao}\ \emph {et~al.}(2003)\citenamefont {Tao},
  \citenamefont {{J.P.~Perdew}}, \citenamefont {{V.N.~Staroverov}},\ and\
  \citenamefont {{G.E.~Scuseria}}}]{TaoPerdewStroverovScuseria:03}%
  \BibitemOpen
  \bibfield  {author} {\bibinfo {author} {\bibfnamefont {J.}~\bibnamefont
  {Tao}}, \bibinfo {author} {\bibnamefont {{J.P.~Perdew}}}, \bibinfo {author}
  {\bibnamefont {{V.N.~Staroverov}}}, \ and\ \bibinfo {author} {\bibnamefont
  {{G.E.~Scuseria}}},\ }\href {\doibase 10.1103/PhysRevLett.91.146401}
  {\bibfield  {journal} {\bibinfo  {journal} {Phys. Rev. Lett.}\ }\textbf
  {\bibinfo {volume} {91}},\ \bibinfo {pages} {146401} (\bibinfo {year}
  {2003})}\BibitemShut {NoStop}%
\bibitem [{\citenamefont {K{\"u}mmel}\ and\ \citenamefont
  {Kronik}(2008)}]{KuemmelKronik:08}%
  \BibitemOpen
  \bibfield  {author} {\bibinfo {author} {\bibfnamefont {S.}~\bibnamefont
  {K{\"u}mmel}}\ and\ \bibinfo {author} {\bibfnamefont {L.}~\bibnamefont
  {Kronik}},\ }\href {\doibase 10.1103/RevModPhys.80.3} {\bibfield  {journal}
  {\bibinfo  {journal} {Rev. Mod. Phys.}\ }\textbf {\bibinfo {volume} {80}},\
  \bibinfo {pages} {3} (\bibinfo {year} {2008})}\BibitemShut {NoStop}%
\bibitem [{\citenamefont {Runge}\ and\ \citenamefont {{E.K.U.
  Gross}}(1984)}]{RungeGross:84}%
  \BibitemOpen
  \bibfield  {author} {\bibinfo {author} {\bibfnamefont {E.}~\bibnamefont
  {Runge}}\ and\ \bibinfo {author} {\bibnamefont {{E.K.U. Gross}}},\ }\href
  {\doibase 10.1103/PhysRevLett.52.997} {\bibfield  {journal} {\bibinfo
  {journal} {Phys. Rev. Lett.}\ }\textbf {\bibinfo {volume} {52}},\ \bibinfo
  {pages} {997} (\bibinfo {year} {1984})}\BibitemShut {NoStop}%
\bibitem [{\citenamefont {Ullrich}(2012)}]{Ullrich:12}%
  \BibitemOpen
  \bibfield  {author} {\bibinfo {author} {\bibfnamefont {C.}~\bibnamefont
  {Ullrich}},\ }\href@noop {} {\emph {\bibinfo {title} {Time-Dependent
  Density-Functional Theory}}}\ (\bibinfo  {publisher} {Oxford University
  Press},\ \bibinfo {address} {Oxford},\ \bibinfo {year} {2012})\BibitemShut
  {NoStop}%
\bibitem [{\citenamefont {Maitra}(2016)}]{Maitra:16}%
  \BibitemOpen
  \bibfield  {author} {\bibinfo {author} {\bibfnamefont {N.~T.}\ \bibnamefont
  {Maitra}},\ }\href {\doibase 10.1063/1.4953039} {\bibfield  {journal}
  {\bibinfo  {journal} {J. Chem. Phys.}\ }\textbf {\bibinfo {volume} {144}},\
  \bibinfo {pages} {220901} (\bibinfo {year} {2016})}\BibitemShut {NoStop}%
\bibitem [{\citenamefont {Onida}\ \emph {et~al.}(2002)\citenamefont {Onida},
  \citenamefont {Reining},\ and\ \citenamefont {Rubio}}]{OnidaReiningRubio:02}%
  \BibitemOpen
  \bibfield  {author} {\bibinfo {author} {\bibfnamefont {G.}~\bibnamefont
  {Onida}}, \bibinfo {author} {\bibfnamefont {L.}~\bibnamefont {Reining}}, \
  and\ \bibinfo {author} {\bibfnamefont {A.}~\bibnamefont {Rubio}},\ }\href
  {\doibase 10.1103/RevModPhys.74.601} {\bibfield  {journal} {\bibinfo
  {journal} {Rev. Mod. Phys.}\ }\textbf {\bibinfo {volume} {74}},\ \bibinfo
  {pages} {601} (\bibinfo {year} {2002})}\BibitemShut {NoStop}%
\bibitem [{\citenamefont {Martin}\ \emph {et~al.}(2016)\citenamefont {Martin},
  \citenamefont {Reining},\ and\ \citenamefont
  {Ceperley}}]{MartinReiningCeperley:16}%
  \BibitemOpen
  \bibfield  {author} {\bibinfo {author} {\bibfnamefont {R.}~\bibnamefont
  {Martin}}, \bibinfo {author} {\bibfnamefont {L.}~\bibnamefont {Reining}}, \
  and\ \bibinfo {author} {\bibfnamefont {D.}~\bibnamefont {Ceperley}},\
  }\href@noop {} {\emph {\bibinfo {title} {Interacting Electrons: Theory and
  Computational Approaches}}}\ (\bibinfo  {publisher} {Cambridge University
  Press},\ \bibinfo {address} {Cambridge},\ \bibinfo {year} {2016})\BibitemShut
  {NoStop}%
\bibitem [{\citenamefont {Capelle}\ and\ \citenamefont
  {Campo~Jr.}(2013)}]{CapelleCampo:13}%
  \BibitemOpen
  \bibfield  {author} {\bibinfo {author} {\bibfnamefont {K.}~\bibnamefont
  {Capelle}}\ and\ \bibinfo {author} {\bibfnamefont {V.~L.}\ \bibnamefont
  {Campo~Jr.}},\ }\href {\doibase 10.1016/j.physrep.2013.03.002} {\bibfield
  {journal} {\bibinfo  {journal} {Phys. Rep.}\ }\textbf {\bibinfo {volume}
  {528}},\ \bibinfo {pages} {91} (\bibinfo {year} {2013})}\BibitemShut
  {NoStop}%
\bibitem [{\citenamefont {Stefanucci}\ and\ \citenamefont
  {Kurth}(2015)}]{StefanucciKurth:15}%
  \BibitemOpen
  \bibfield  {author} {\bibinfo {author} {\bibfnamefont {G.}~\bibnamefont
  {Stefanucci}}\ and\ \bibinfo {author} {\bibfnamefont {S.}~\bibnamefont
  {Kurth}},\ }\href {\doibase 10.1021/acs.nanolett.5b03294} {\bibfield
  {journal} {\bibinfo  {journal} {Nano~Lett.}\ }\textbf {\bibinfo {volume}
  {15}},\ \bibinfo {pages} {8020} (\bibinfo {year} {2015})}\BibitemShut
  {NoStop}%
\bibitem [{\citenamefont {Kurth}\ and\ \citenamefont
  {Stefanucci}(2017)}]{KurthStefanucci:17}%
  \BibitemOpen
  \bibfield  {author} {\bibinfo {author} {\bibfnamefont {S.}~\bibnamefont
  {Kurth}}\ and\ \bibinfo {author} {\bibfnamefont {G.}~\bibnamefont
  {Stefanucci}},\ }\href {http://stacks.iop.org/0953-8984/29/i=41/a=413002}
  {\bibfield  {journal} {\bibinfo  {journal} {J. Phys.: Condens. Matter}\
  }\textbf {\bibinfo {volume} {29}},\ \bibinfo {pages} {413002} (\bibinfo
  {year} {2017})}\BibitemShut {NoStop}%
\bibitem [{\citenamefont {Stefanucci}\ and\ \citenamefont {van
  Leeuwen}(2013)}]{svl-book}%
  \BibitemOpen
  \bibfield  {author} {\bibinfo {author} {\bibfnamefont {G.}~\bibnamefont
  {Stefanucci}}\ and\ \bibinfo {author} {\bibfnamefont {R.}~\bibnamefont {van
  Leeuwen}},\ }\href@noop {} {\emph {\bibinfo {title} {Nonequilibrium Many-Body
  Theory of Quantum Systems: A Modern Introduction}}}\ (\bibinfo  {publisher}
  {Cambridge University Press},\ \bibinfo {address} {Cambridge},\ \bibinfo
  {year} {2013})\BibitemShut {NoStop}%
\bibitem [{Note1()}]{Note1}%
  \BibitemOpen
  \bibinfo {note} {The non-interacting retarded and advanced GFs $\protect \bm
  {G}_0^{r,a}=(\omega -\protect \bm {H}_0-\protect \bm {\Sigma
  }^{r,a}_{\protect \rm T}-\protect \bm {\Sigma }^{r,a}_{\protect \rm R})
  ^{-1}$ are independent of the bias by construction, while the lesser GF is in
  principle bias-dependent via the tip Fermi function $f_{\protect \rm
  T}=f(\omega -V)$: $G_0^<=G_0^r(f_{\protect \rm T}{\protect \bm {\Gamma
  }}_{\protect \rm T}+f_{\protect \rm R}{\protect \bm {\Gamma }}_{\protect \rm
  R})G_0^a$. However, in the limit ${\protect \bm {\Gamma }}_{\protect \rm
  T}\rightarrow 0$ the bias dependence vanishes as $\partial {G}^<_0/\partial
  {V}\sim {\protect \bm {\Gamma }}_{\protect \rm T}$. As the interacting GFs
  can be written diagrammatically in terms of the non-interacting GFs and the
  electron-electron interaction does not depend on the bias either, it follows
  that the interacting GFs must also be independent of the bias in the limit
  ${\protect \bm {\Gamma }}_{\protect \rm T}\rightarrow 0$, i.e. $\partial
  {G}^{r,a,<}/\partial {V}\rightarrow 0$.}\BibitemShut {Stop}%
\bibitem [{\citenamefont {Meir}\ and\ \citenamefont
  {Wingreen}(1992)}]{Meir:PRL:1992}%
  \BibitemOpen
  \bibfield  {author} {\bibinfo {author} {\bibfnamefont {Y.}~\bibnamefont
  {Meir}}\ and\ \bibinfo {author} {\bibfnamefont {N.~S.}\ \bibnamefont
  {Wingreen}},\ }\href {\doibase 10.1103/PhysRevLett.68.2512} {\bibfield
  {journal} {\bibinfo  {journal} {Phys. Rev. Lett}\ }\textbf {\bibinfo {volume}
  {68}},\ \bibinfo {pages} {2512} (\bibinfo {year} {1992})}\BibitemShut
  {NoStop}%
\bibitem [{Note2()}]{Note2}%
  \BibitemOpen
  \bibinfo {note} {Transformation from symmetric voltage drop $V_{\protect \rm
  T}=-V_{\protect \rm R}=V/2$ to a completely asymmetric voltage drop,
  $V_{\protect \rm T}=V$ and $V_{\protect \rm R}=0$, is achieved by a spatially
  constant shift of the gate potential by $-V/2$, and subsequent substitution
  of the integration variable, $\omega \rightarrow \omega +(V+V_{\protect \rm
  xc})/2$.}\BibitemShut {Stop}%
\bibitem [{\citenamefont {Anderson}(1961)}]{Anderson:61}%
  \BibitemOpen
  \bibfield  {author} {\bibinfo {author} {\bibfnamefont {P.~W.}\ \bibnamefont
  {Anderson}},\ }\href {\doibase 10.1103/PhysRev.124.41} {\bibfield  {journal}
  {\bibinfo  {journal} {Phys. Rev.}\ }\textbf {\bibinfo {volume} {124}},\
  \bibinfo {pages} {41} (\bibinfo {year} {1961})}\BibitemShut {NoStop}%
\bibitem [{\citenamefont {Kurth}\ and\ \citenamefont
  {Stefanucci}(2016)}]{KurthStefanucci:16}%
  \BibitemOpen
  \bibfield  {author} {\bibinfo {author} {\bibfnamefont {S.}~\bibnamefont
  {Kurth}}\ and\ \bibinfo {author} {\bibfnamefont {G.}~\bibnamefont
  {Stefanucci}},\ }\href {\doibase 10.1103/PhysRevB.94.241103} {\bibfield
  {journal} {\bibinfo  {journal} {Phys. Rev. B}\ }\textbf {\bibinfo {volume}
  {94}},\ \bibinfo {pages} {241103(R)} (\bibinfo {year} {2016})}\BibitemShut
  {NoStop}%
\bibitem [{\citenamefont {Perdew}\ \emph {et~al.}(1982)\citenamefont {Perdew},
  \citenamefont {Parr}, \citenamefont {Levy},\ and\ \citenamefont
  {Balduz}}]{PerdewParrLevyBalduz:82}%
  \BibitemOpen
  \bibfield  {author} {\bibinfo {author} {\bibfnamefont {J.~P.}\ \bibnamefont
  {Perdew}}, \bibinfo {author} {\bibfnamefont {R.}~\bibnamefont {Parr}},
  \bibinfo {author} {\bibfnamefont {M.}~\bibnamefont {Levy}}, \ and\ \bibinfo
  {author} {\bibfnamefont {J.~L.}\ \bibnamefont {Balduz}},\ }\href {\doibase
  10.1103/PhysRevLett.49.1691} {\bibfield  {journal} {\bibinfo  {journal}
  {Phys. Rev. Lett.}\ }\textbf {\bibinfo {volume} {49}},\ \bibinfo {pages}
  {1691} (\bibinfo {year} {1982})}\BibitemShut {NoStop}%
\bibitem [{\citenamefont {Stefanucci}\ and\ \citenamefont
  {Kurth}(2013)}]{StefanucciKurth:13}%
  \BibitemOpen
  \bibfield  {author} {\bibinfo {author} {\bibfnamefont {G.}~\bibnamefont
  {Stefanucci}}\ and\ \bibinfo {author} {\bibfnamefont {S.}~\bibnamefont
  {Kurth}},\ }\href {\doibase 10.1002/pssb.201349181} {\bibfield  {journal}
  {\bibinfo  {journal} {Phys.~Stat.~Sol.~(b)}\ }\textbf {\bibinfo {volume}
  {250}},\ \bibinfo {pages} {2378} (\bibinfo {year} {2013})}\BibitemShut
  {NoStop}%
\bibitem [{\citenamefont {Bergfield}\ \emph {et~al.}(2012)\citenamefont
  {Bergfield}, \citenamefont {Liu}, \citenamefont {Burke},\ and\ \citenamefont
  {Stafford}}]{blbs.2012}%
  \BibitemOpen
  \bibfield  {author} {\bibinfo {author} {\bibfnamefont {J.~P.}\ \bibnamefont
  {Bergfield}}, \bibinfo {author} {\bibfnamefont {Z.-F.}\ \bibnamefont {Liu}},
  \bibinfo {author} {\bibfnamefont {K.}~\bibnamefont {Burke}}, \ and\ \bibinfo
  {author} {\bibfnamefont {C.~A.}\ \bibnamefont {Stafford}},\ }\href {\doibase
  10.1103/PhysRevLett.108.066801} {\bibfield  {journal} {\bibinfo  {journal}
  {Phys. Rev. Lett.}\ }\textbf {\bibinfo {volume} {108}},\ \bibinfo {pages}
  {066801} (\bibinfo {year} {2012})}\BibitemShut {NoStop}%
\bibitem [{\citenamefont {Beenakker}(1991)}]{rate-paper1}%
  \BibitemOpen
  \bibfield  {author} {\bibinfo {author} {\bibfnamefont {C.~W.~J.}\
  \bibnamefont {Beenakker}},\ }\href {\doibase 10.1103/PhysRevB.44.1646}
  {\bibfield  {journal} {\bibinfo  {journal} {Phys. Rev. B}\ }\textbf {\bibinfo
  {volume} {44}},\ \bibinfo {pages} {1646} (\bibinfo {year}
  {1991})}\BibitemShut {NoStop}%
\bibitem [{\citenamefont {Stefanucci}\ and\ \citenamefont
  {Kurth}(2011)}]{sk.2011}%
  \BibitemOpen
  \bibfield  {author} {\bibinfo {author} {\bibfnamefont {G.}~\bibnamefont
  {Stefanucci}}\ and\ \bibinfo {author} {\bibfnamefont {S.}~\bibnamefont
  {Kurth}},\ }\href {\doibase 10.1103/PhysRevLett.107.216401} {\bibfield
  {journal} {\bibinfo  {journal} {Phys. Rev. Lett.}\ }\textbf {\bibinfo
  {volume} {107}},\ \bibinfo {pages} {216401} (\bibinfo {year}
  {2011})}\BibitemShut {NoStop}%
\bibitem [{\citenamefont {Tr\"oster}\ \emph {et~al.}(2012)\citenamefont
  {Tr\"oster}, \citenamefont {Schmitteckert},\ and\ \citenamefont
  {Evers}}]{tse.2012}%
  \BibitemOpen
  \bibfield  {author} {\bibinfo {author} {\bibfnamefont {P.}~\bibnamefont
  {Tr\"oster}}, \bibinfo {author} {\bibfnamefont {P.}~\bibnamefont
  {Schmitteckert}}, \ and\ \bibinfo {author} {\bibfnamefont {F.}~\bibnamefont
  {Evers}},\ }\href {\doibase 10.1103/PhysRevB.85.115409} {\bibfield  {journal}
  {\bibinfo  {journal} {Phys. Rev. B}\ }\textbf {\bibinfo {volume} {85}},\
  \bibinfo {pages} {115409} (\bibinfo {year} {2012})}\BibitemShut {NoStop}%
\bibitem [{\citenamefont {Bulla}\ \emph {et~al.}(2008)\citenamefont {Bulla},
  \citenamefont {Costi},\ and\ \citenamefont {Pruschke}}]{Bulla:RMP:2008}%
  \BibitemOpen
  \bibfield  {author} {\bibinfo {author} {\bibfnamefont {R.}~\bibnamefont
  {Bulla}}, \bibinfo {author} {\bibfnamefont {T.~A.}\ \bibnamefont {Costi}}, \
  and\ \bibinfo {author} {\bibfnamefont {T.}~\bibnamefont {Pruschke}},\ }\href
  {\doibase 10.1103/RevModPhys.80.395} {\bibfield  {journal} {\bibinfo
  {journal} {Rev. Mod. Phys.}\ }\textbf {\bibinfo {volume} {80}},\ \bibinfo
  {pages} {3950} (\bibinfo {year} {2008})}\BibitemShut {NoStop}%
\bibitem [{\citenamefont {Motahari}\ \emph {et~al.}(2016)\citenamefont
  {Motahari}, \citenamefont {Requist},\ and\ \citenamefont
  {Jacob}}]{Motahari:PRB:2016}%
  \BibitemOpen
  \bibfield  {author} {\bibinfo {author} {\bibfnamefont {S.}~\bibnamefont
  {Motahari}}, \bibinfo {author} {\bibfnamefont {R.}~\bibnamefont {Requist}}, \
  and\ \bibinfo {author} {\bibfnamefont {D.}~\bibnamefont {Jacob}},\ }\href
  {\doibase 10.1103/PhysRevB.94.235133} {\bibfield  {journal} {\bibinfo
  {journal} {Phys. Rev. B}\ }\textbf {\bibinfo {volume} {94}},\ \bibinfo
  {pages} {235133} (\bibinfo {year} {2016})}\BibitemShut {NoStop}%
\bibitem [{\citenamefont {Kraisler}\ and\ \citenamefont
  {Kronik}(2013)}]{KraislerKronik:13}%
  \BibitemOpen
  \bibfield  {author} {\bibinfo {author} {\bibfnamefont {E.}~\bibnamefont
  {Kraisler}}\ and\ \bibinfo {author} {\bibfnamefont {L.}~\bibnamefont
  {Kronik}},\ }\href {\doibase 10.1103/PhysRevLett.110.126403} {\bibfield
  {journal} {\bibinfo  {journal} {Phys. Rev. Lett.}\ }\textbf {\bibinfo
  {volume} {110}},\ \bibinfo {pages} {126403} (\bibinfo {year}
  {2013})}\BibitemShut {NoStop}%
\bibitem [{\citenamefont {Baer}\ and\ \citenamefont
  {Neuhauser}(2005)}]{BaerNeuhauser:05}%
  \BibitemOpen
  \bibfield  {author} {\bibinfo {author} {\bibfnamefont {R.}~\bibnamefont
  {Baer}}\ and\ \bibinfo {author} {\bibfnamefont {D.}~\bibnamefont
  {Neuhauser}},\ }\href {\doibase 10.1103/PhysRevLett.94.043002} {\bibfield
  {journal} {\bibinfo  {journal} {Phys. Rev. Lett.}\ }\textbf {\bibinfo
  {volume} {94}},\ \bibinfo {pages} {043002} (\bibinfo {year}
  {2005})}\BibitemShut {NoStop}%
\bibitem [{\citenamefont {Yamada}\ \emph {et~al.}(2016)\citenamefont {Yamada},
  \citenamefont {Feng}, \citenamefont {Hoskins}, \citenamefont {Fenk},\ and\
  \citenamefont {Dunietz}}]{YamadaFengHoskinsFenkDunietz:16}%
  \BibitemOpen
  \bibfield  {author} {\bibinfo {author} {\bibfnamefont {A.}~\bibnamefont
  {Yamada}}, \bibinfo {author} {\bibfnamefont {Q.}~\bibnamefont {Feng}},
  \bibinfo {author} {\bibfnamefont {A.}~\bibnamefont {Hoskins}}, \bibinfo
  {author} {\bibfnamefont {K.~D.}\ \bibnamefont {Fenk}}, \ and\ \bibinfo
  {author} {\bibfnamefont {B.~D.}\ \bibnamefont {Dunietz}},\ }\href {\doibase
  10.1021/acs.nanolett.6b02241} {\bibfield  {journal} {\bibinfo  {journal}
  {Nano Lett.}\ }\textbf {\bibinfo {volume} {16}},\ \bibinfo {pages} {6092}
  (\bibinfo {year} {2016})},\ \Eprint
  {http://arxiv.org/abs/https://doi.org/10.1021/acs.nanolett.6b02241}
  {https://doi.org/10.1021/acs.nanolett.6b02241} \BibitemShut {NoStop}%
\end{thebibliography}%

\end{document}